\documentclass[aps,pra,twocolumn,amsfonts, amssymb, amsmath, amsthm,groupedaddress]{revtex4-1}
\usepackage{url}
\usepackage{color}
\usepackage{graphicx}
\usepackage{amsthm}

\newcommand{\bra}[1]{{\langle#1|}}
\newcommand{\ket}[1]{{|#1\rangle}}
\DeclareMathOperator{\Tr}{\operatorname{Tr}}

\begin{document}

\title{Quantum Steganography over Noiseless Channels: Achievability and Bounds}


\author{Chris Sutherland$^1$ and Todd A. Brun$^{1,2}$}
\affiliation{$^1$Department of Physics,
$^2$Ming Hsieh Department of Electrical Engineering, \\
University of Southern California, Los Angeles, California}


\date{\today}

\begin{abstract}
Quantum steganography is the study of hiding secret quantum information by encoding it into what an eavesdropper would perceive as an innocent-looking message. Here we study an explicit steganographic encoding for Alice to hide her secret message in the syndromes of an error-correcting code, so that the encoding simulates a given noisy quantum channel.  We calculate achievable rates of steganographic communication over noiseless quantum channels using this encoding. We give definitions of secrecy and reliability for the communication process, and with these assumptions derive upper bounds on the amount of steganographic communication possible, and show that these bounds match the communication rates achieved with our encoding.  This gives a steganographic capacity for a noiseless channel emulating a given noisy channel.
\end{abstract}


\maketitle

\section{Introduction}\label{intro}

The study of steganography is perhaps best motivated by considering an example. Suppose two political protestors Alice and Bob are arrested and put into two widely separated jail cells. The warden allows them to communicate with hand-written letters that he reads before delivering. However, if the warden reads anything in the letters that he finds suspicious (such as a possible escape plan), then he will not deliver the letter. Luckily, Alice and Bob exchanged a secret key before their incarceration. Can Alice and Bob communicate their escape plan to each other without arousing the warden's suspicions? This is where the study of steganography comes into play.

The science of sending information through seemingly innocuous messages has a long history, dating back to at least 440 B.C.\cite{greekstego}. It is worth making clear its differences from cryptography. In cryptography, a secret message (the {\it plaintext}) is encrypted using the shared secret key, and the resulting {\it ciphertext} is then sent to the desired receiver to be decoded. If an eavesdropper (Eve) observes the ciphertext, she cannot decode it without the secret key. However, she will know that there {\it is} a secret message, since Alice is sending apparent gibberish to Bob.

By contrast, if Alice uses a steganographic encoding, she hides the secret message (or {\it stegotext}) into a larger {\it covertext}, which appears to Eve as an innocuous message.  The hidden message may or may not be encrypted itself, but the main line of defense is that the eavesdropper is unaware that a message is even being sent.

During WWII, a Japanese spy named Velvalee Dickinson sent classified information to neutral South America. She was a dealer in dolls, and her letters discussed the quantity and type of doll to ship. The covertext was the doll orders, while the concealed stegotext was encoded information about battleship movements \cite{ww2}.

The quantum analogue of cryptography has been widely studied \cite{norbertcrypto}. However, the quantum analogue of steganography is still in a relatively early stage. There have been a number of different proposals for encoding quantum information steganographically, or encoding classical information into quantum states or channels \cite{natoristego,banerjeestego}. In this paper we consider hiding secret messages as error syndromes of a quantum error-correcting code \cite{gea2002hiding}. This approach to quantum steganography has been studied in detail by Shaw and Brun, with explicit encoding and decoding procedures and calculated rates of communication and secret key consumption \cite{shaw2011quantum,shaw2010hiding}. It was shown that such schemes can hide both quantum and classical information, with a quantitative measure of secrecy, even in the presence of a noisy physical channel.  When the error rate of the physical channel is lower than the eavesdropper's expectation, it is possible to achieve non-zero asymptotic rates of communication.  (If the eavesdropper has exact knowledge of the channel, secret communication may still be possible, but the amount of secret information that can be transmitted in general grows sublinearly with the number of channel uses.)

More recently, a closely related idea has been studied under the name of  quantum covert communication  \cite{qcovert1,qcovert2,qcovert3,qcovert4,qcovert5}. Many of the ideas in this paper are closely related to steganographic requirements, such as secrecy and recoverability. This is not surprising, since covert quantum communication can be seen as a special case of quantum steganography over noisy quantum channels in the case when the eavesdropper has exact knowledge of the channel, and where Eve assumes the channel is idle (so only noise is being transmitted). Similarly, quantum steganography is a type of covert quantum communication where Eve knows about the covertext communication but not the hidden stegotext, and where Eve may not have perfect knowledge of the channel.  The work on covert communication has generally found that, if Eve has exact knowledge of the channel, the amount of secret communication that can be done grows like the square root of the number of channel used.

The goal of this paper is to formalize the assumptions and reasonable conditions of quantum steganography introduced in \cite{shaw2011quantum}, and to give upper bounds on the achievable rates of quantum communication while remaining secure from an eavesdropper's suspicion, for the special case when the true underlying channel is noiseless.  Our results include achievability results as well as converse proofs for quantum steganography.

In Section \ref{qstego} we formalize our notion of quantum steganography where secret messages are hidden in the syndromes of an error-correcting code, and outline a specific steganographic encoding where Alice is able to emulate any general quantum channel $\mathcal{N}$ on her encoded secret message and covertext.  We work out specific examples for the bit-flip channel and the depolarizing channel, before giving the more general result. In Section \ref{noiseless} we prove upper bounds on the amount of steganographic communication possible, and show that these bounds are asymptotically equal to the rates achieved in the previous section.

The assumption that the physical channel is noiseless greatly simplifies the analysis.  However, we believe that the main intuition underlying this approach will apply equally well in the case of a noisy channel. We will end this paper with a discussion of how to extend this work to the case where the physical channel between the two parties is noisy.

\section{Quantum Steganography:  Achievability}\label{qstego}

As discussed in the introduction, there have been several approaches to generalizing steganography to the quantum setting. Here we will make explicit the notion of quantum steganography based on syndromes of quantum error-correcting codes. We assume that Eve expects to see quantum information passing through a noisy quantum channel. However, the actual physical channel is assumed to be noiseless.  This is obviously an idealized assumption, which greatly simplifies the analysis; we will discuss below how it might be justified at least as an approximation.

Alice wants send a secret message steganographically to Bob.  Using her shared secret key, she encodes the stego text into a codeword of a quantum error-correcting code (QECC) with errors applied to it, and sends it to Bob.  The codeword encodes an innocent state; the stego text is conveyed in the errors.  If Eve were to perform measurements on this codeword, it would be indistinguishable from an innocent encoded covertext that had passed through a given noisy quantum channel to Bob.

Before discussing how to quantify the security of a quantum steganographic protocol, let us make clear what Alice is trying to achieve. Alice wants to encode an innocent covertext state, together with her secret message, into an $N$-qubit codeword in such a way that it cannot be distinguished from the covertext alone encoded into a quantum error-correcting code that has undergone typical errors induced by the quantum channel $\mathcal{N}^{\otimes N}$.  The steganographic encoding works by mapping all possible secret messages onto syndromes of the QECC.  This encoding is not limited to classical messages:  it is possible to encode a quantum state by preparing the codeword in a superposition of different error syndromes.

In analyzing this quantum steganography protocol, we make the following assumptions.  Alice is communicating with Bob by a quantum channel that is actually noiseless.  But the eavesdropper, Eve, believes that this channel is noisy, perhaps because Alice and Bob have been systematically making the channel appear noisier than it actually is.  Because Alice and Bob have been systematically deceiving Eve in this way, we assume that they know (at least fairly closely) what Eve's estimate of the channel is.  Before the protocol began, Alice and Bob shared with each other a secret key:  an arbitrarily long string of random bits.  This key is known only to the two of them.  But once the protocol begins, they cannot communicate except through channels that can be monitored by Eve.  Alice sends an innocent-looking message to Bob over the channel.  This is a covertext state $\rho_c$, encoded into an error-correcting code; it is assumed that the choice of code is known to Eve, and this code should be a plausible choice for the noisy channel that Eve believes exists.

One important caveat for this section:  we will be considering the case where the QECC that Alice uses is nondegenerate. That is, each typical error corresponds to a unique error syndrome.  This allows Alice to communicate as much steganographic information as possible, and it allows us to ignore the details of which QECC is being used.  Methods similar to those in this section should also work for degenerate codes; but in that case, the encoding will be strongly dependent on the properties of the particular code, since the typical errors must first be grouped into equivalent sets, and then the possible messages mapped into these sets.  We also use this assumption in the next section to get specific expressions for the upper bound on the secret communication rate.

To clarify how the encoding works, we start by considering two examples for relatively simple channels: first, the case where Alice is emulating a bit flip channel $\mathcal{N}^{BF}_{p}$ on the codeword, and second, the case where she is emulating the depolarizing channel.  Finally we consider a more general error map $\mathcal{N}^{\otimes N}$. The message qubits are encoded into into the error syndromes of the codeword of the QECC she is using.

\subsection{The Bit Flip Channel}

Suppose that Eve believes the channel connecting Alice and Bob to be a bit flip channel, with a probability $p$ of error per qubit sent.  (The actual physical channel is noiseless, as assumed above.)  Alice sends a codeword of length $N$ to Bob, encoding some ``innocent'' covertext state $\rho_c$.  The errors in the codewords that Alice sends to Bob should be binomially distributed:  $pN$ is the mean number of errors of this distribution, and the variance is $(1-p)pN$. The total probability that there is an error of weight $w$ on the codeword should be
\begin{equation}
p_{k}={N \choose w}p^{w}(1-p)^{N-w} .
\label{eq:binomial_dist}
\end{equation}
There are
\[
{N \choose w} \equiv \frac{N!}{w! (N-w)!}
\]
such errors, all with equal probability $p^w (1-p)^{N-w}$.

If $N$ is large, then it is extremely likely that the number of bit flips will be a {\it typical} error---that is, an error of weight $w$ within a narrow range about the mean $pN$.  Alice's encoding will make use of these typical errors.  For each $w$ from $Np(1-\delta)$ to $Np(1+\delta)$, where $\sqrt{(1-p)/pN} \ll \delta \ll 1$, Alice chooses at random a set of $C_w$ possible error strings of weight $w$.  (An {\it error string} of weight $w$ is a string of $N$ bits, with a 1 at every location with a bit flip and 0 at every location with no error.)  This random choice is made using the shared secret key with Bob, so that Bob also knows which set of errors is being used to encode secret messages, but Eve (who does not share the key) could not know this.

Let these sets of error strings of weight $w$ be called $\{S_{w}\}$, and the set of all strings used in the encoding is
\begin{equation}
S=\bigcup\limits_{w}S_{w}.
\end{equation}
We sum up
\begin{equation}
C=\sum_{w=Np(1-\delta)}^{Np(1+\delta)} C_{w} = |S| .
\end{equation}
So the total number of strings in the set $S$ is $C$.  This number $C$ is the total number of possible distinct secret messages that Alice can send to Bob (though she may also send {\it superpositions} of these messages).  We assume all these messages to be equally likely.  So the message encodes $M = \log_2 C$ bits (or qubits) of information.

Define the probability $q=1/C$. These error strings $S$ are typical strings (using the definition of weak typicality from information theory).  Eve should not be suspicious at seeing such an error string, since it matches a probable result for the channel that she expects.  For this encoding to be indistinguishable from the bit flip channel, the probability of the message being an error string of weight $w$ should equal the value from the distribution in Eq.~(\ref{eq:binomial_dist}) above.  This means we want to satisfy
\begin{equation}
qC_{w} = \frac{C_{w}}{C} = p_{w} .
\end{equation}
Clearly we must have
\[
C_{w}\leq {N \choose w} ,
\]
for all $w$ in the typical range. This implies that:
\begin{align}
C_{w}p^{w}(1-p)^{N-w} \leq {N \choose w}p^{w}(1-p)^{N-w} &= C_{w} q , \nonumber \\
\Rightarrow p^{w}(1-p)^{N-w} &\leq q .
\label{eq:bitflipconstraint}
\end{align}
To communicate the maximum amount of information steganographically we want $C$ to be as large as possible, which means we want $q$ to be as small as possible. The constraint in Eq.~(\ref{eq:bitflipconstraint}) then gives us
\begin{equation}
q=p^{Np(1-\delta)}(1-p)^{N(1-p+p\delta)} .
\end{equation}
So Alice can send $M$ stego qubits to Bob, where
\begin{align}
M =& \log_{2}C = \log_{2}1/q \nonumber \\
=& N(-p\log_{2}p-(1-p)\log_{2}(1-p) \nonumber \\
&+\delta(p\log_{2}p-p\log_{2}(1-p))) \nonumber \\
=& N(h(p)-\delta p\log_{2}((1-p)/p)) \nonumber \\
\approx& Nh(p) ,
\end{align}
where $h(p)=-p\log_{2}p-(1-p)\log_{2}(1-p)$ is the entropy of the bit flip channel on one qubit. Therefore, with this encoding Alice can send almost $Nh(p)$ bits.

In \cite{shaw2011quantum} it is shown that the diamond norm distance between the channel $(\mathcal{N}_{p}^{BF})^{\otimes N}$ and Alice's encoding is exponentially small in $N$.  This justifies the claim that this protocol will not arouse suspicion from Eve. In section III we use a slightly modified definition of secrecy that allows us to prove the converse bound on this rate of stego communication by information theoretic techniques.  That means that this encoding is essentially optimal:  the maximum rate of steganographic communication for a nondegenerate code in the case of the bit flip channel is $h(p)$.

\subsection{Depolarizing Channel}

Here we will consider the scenario where the channel Alice is emulating is the depolarizing channel. It turns out that due to the symmetric nature of the depolarizing channel the encoding looks quite similar to that of the bit flip channel. Recall that the depolarizing channel acting on a single qubit $\rho$ is given by
\begin{equation*}
\mathcal{N}^{DC}_{p}(\rho) = (1-p)\rho + (p/3)(X\rho X+Y\rho Y+Z \rho Z).
\end{equation*}
Applying this channel on $N$ qubits, the total probability of all errors with exactly $n_{1}$ $X$, $n_{2}$ $Y$, and $n_{3}$ $Z$ errors (and $n_{4} = N - n_1 - n_2 - n_3$ identity ``errors'') is
\begin{equation*}
p(n_{1},n_{2},n_{3},n_{4})=\frac{N!}{n_{1}!n_{2}!n_{3}!n_{4}!}(p/3)^{n_1+n_2+n_3}(1-p)^{n_4}.
\end{equation*}
Notice that instead of specifying $n_{1}, n_{2},$ and $n_{3}$ exactly, we can instead talk about errors with weight $w=n_{1}+n_{2}+n_{3}$. It follows by simple calculation that the total probability of all errors of weight $w$ is
\begin{equation*}
p(w)=3^w{N\choose w}(p/3)^w(1-p)^{N-w}={N \choose w}p^w(1-p)^{N-w},
\end{equation*}
which is just a binomial distribution in $w$. As in the bit flip case, we will need say what strings of errors are typical. There are a number of ways we could specify this, but for simplicity we will consider weights $w$ that lie between $Np(1-\delta)$ and $Np(1+\delta)$ for $\sqrt{(1-p)/pN}\ll\delta\ll 1$. The astute reader will notice that this set includes some errors that are not typical: for instance, it includes errors of weight $w$ where all (or most) of the errors are $X$'s and none (or few) are $Y$'s or $Z$'s. If such errors are used as codewords, they might make Eve suspicious. Still, the effect of this is not too large, because this set is still dominated by typical errors, and the probabilities of these strings are similar to the expected probabilities of atypical errors. With this definition of typicality, we can follow the exact same encoding given for the bit flip code using errors with weight $w$, except that the set of errors of weight $w$ is now of size
\[
\left(\begin{array}{c} N \\ w \end{array}\right) 3^w ,
\]
and errors of weight $w$ have probability $(p/3)^w (1-p)^{N-w}$.  This then leads to the following encoding rate:
\begin{align}
M &= N(-p\log_{2}(p/3)-(1-p)\log_{2}(1-p) \nonumber \\
& +\delta(p\log_{2}(p/3)-p\log_{2}(1-p))) \nonumber \\
&= N(s(p)+\delta(p\log_{2}(p/3)-p\log_{2}(1-p)) \nonumber \\ 
&\approx Ns(p)
\end{align}
where we have defined $s(p)=-p\log_{2}(p/3)-(1-p)\log_{2}(1-p)$ to be the entropy of the depolarizing channel on one qubit.

\subsection{General Channels}

\subsubsection{Special case:  random unitaries}

Consider a quantum channel acting on a single qubit of the form
\begin{equation}\label{eq:randomUnitaryChannel}
\mathcal{N}(\rho) = \sum_{i=1}^k p_{i} U_{i} \rho U_{i}^{\dagger} ,
\end{equation}
where the operators $U_i$ are all unitary, so $U_i U_i^\dagger = U_i^\dagger U_i = I$.  The set of Kraus operators $\{\sqrt{p_i} U_i\}$ can be thought of as a set of possible single-qubit unitary errors $U_i$ that occur with probability $p_i$.  Note that both the bit-flip and depolarizing channels are special cases of the random unitary channel, as is any Pauli channel.  The channel acts on an $N$-qubit encoded state $\rho$ as $\mathcal{N}^{\otimes N}(\rho)$.

The total probability of all errors with $n_{1}$ $U_{1}$ errors, $n_{2}$ $U_{2}$ errors, and so forth, is given by the multinomial distribution:
\begin{equation}
p(n_{1},\ldots,n_{k})=\frac{N!}{n_{1}!\cdots n_{k}!}p_{1}^{n_{1}}\cdots p_{k}^{n_{k}}.
\end{equation}
Now consider weights $n_{j}$ in the range from $Np_{j}(1-\delta)$ to $Np_{j}(1+\delta)$, where $\delta$ is large enough that this set includes all the typical strings. (This definition can be modified, but for simplicity we stick with it in this paper.)  Randomly choose $C_{n_{1},\ldots,n_{k}}$ strings with weights $n_1,n_2,\ldots,n_k$ in this range such that $n_{1}+\ldots+n_{k}=N$. As with the bit flip and depolarizing channels, let these sets of strings be called $S_{n_{1},\ldots,n_{k}}$ and let $S$ denote the union of all these sets of strings, which are a subset of the typical strings. For all weights $n_1,\ldots,n_k$ outside the typical set, we let $C_{n_{1},\ldots,n_{k}}=0$.  The total number of strings in the set $S$ is $C$:
\begin{equation}
C = \sum_{n_1,\ldots,n_k} C_{n_{1},\ldots,n_{k}} .
\end{equation}
Defining $q \equiv 1/C$, we want to satisfy 
\begin{equation}
C_{n_{1},\ldots,n_{k}}q=C_{n_{1},\ldots,n_{k}}/C=p(n_{1},\ldots,n_{k})
\end{equation}
for all weights $n_1,\ldots,n_k$ in the typical set, so that Eve does not become suspicious. Also, clearly $C_{n_{1},\ldots,n_{k}}$ must be less than $\frac{N!}{n_{1}!\cdots n_{k}!}$. This implies that:
\begin{align}
&C_{n_{1},\ldots,n_{k}}p_{1}^{n_{1}}\cdots p_{k}^{n_{k}}\leq\frac{N!}{n_{1}!\cdots n_{k}!}p_{1}^{n_{1}}\cdots p_{k}^{n_{k}} \nonumber \\
&C_{n_{1},\ldots,n_{k}}p_{1}^{n_{1}}\cdots p_{k}^{n_{k}}\leq C_{n_{1},\ldots,n_{k}}q \nonumber \\
&p_{1}^{n_{1}}\cdots p_{k}^{n_{k}}\leq q.
\end{align}
Notice that this time we cannot simply plug in the lower bounds of the sums for $n_{j}$, as we did for the depolarizing and bit flip channels, because we have the additional constraint that $n_{1}+\ldots+n_{k}=N$. However, the same general argument applies.  Inside the set of typical weights, there is a string $\tilde{n}_1,\cdots,\tilde{n}_k$ with $|\tilde{n}_j/N - p_j| \le \delta p_j$ for all $j$, that maximizes the probability:
\begin{equation}
p_{\rm max} \equiv p_1^{\tilde{n}_1} p_2^{\tilde{n}_2} \cdots p_k^{\tilde{n}_k} .
\end{equation}
We can choose $q=p_{\rm max}$, and use this to put a bound on the number of stego qubits M Alice can send to Bob:
\begin{align}
M&=\log_{2}C =-\log_{2}(q) = -\log_2 p_{\rm max} \nonumber\\
&= -\tilde{n}_{1}\log_{2}(p_{1})-\ldots-\tilde{n}_{k}\log_{2}(p_{k}) \nonumber \\
&= N(-\frac{\tilde{n}_{1}}{N}\log_{2}(p_{1})-\ldots-\frac{\tilde{n}_{k}}{N}\log_{2}(p_{k})) \nonumber \\
&\ge N(1-\delta)(-\sum_{i=1}^{k}p_{i}\log_{2}(p_{i})) \nonumber\\
&= N(1-\delta)H(p_1,\ldots,p_k) .
\label{eq:randomUnitariesRate}
\end{align}
So in the limit of large $N$, we should approach a rate $H(p_1,\ldots,p_k)$ with this encoding.

\subsubsection{Encoding general channels across multiple code blocks}

This argument does not necessarily apply directly to a general quantum channel, since the probabilities of the different outcomes can be state dependent.  However, we should be able to do a similar type of encoding for a general quantum channel $\mathcal{N}$ by encoding across multiple code blocks. Consider a general quantum channel acting on a single qubit as
\begin{equation}\label{eq:onequbitchannel}
\mathcal{N}(\rho) = \sum_{i=1}^k A_{i} \rho A_{i}^{\dagger} .
\end{equation}
The channel acts on an $N$-qubit encoded state $\rho$ as $\mathcal{N}^{\otimes N}(\rho)$, where we will let $N$ become large.  For most states $\rho$, we can well approximate this $N$-qubit channel by a sum over the {\it typical} errors \cite{Klesse07,Klesse08},
\begin{equation}\label{eq:Nqubitchannel}
\mathcal{N}^{\otimes N}(\rho) \approx \sum_{\underline{i}\in \mathcal T} E_{\underline{i}} \rho E^\dagger_{\underline{i}},
\end{equation}
where $\rho$ is now the $N$-qubit codeword, the index is $\underline{i} = i_1i_2\ldots i_N$, the typical error $E_{\underline{i}}$ is
\begin{equation}
E_{\underline{i}} = A_{i_1} \otimes A_{i_2} \otimes \cdots \otimes A_{i_N} ,
\end{equation}
and $\mathcal{T}$ is the set of typical sequences $\underline{i}$ \cite{wilde2013quantum}.

We assume that the QECC Alice uses is one that can correct the typical errors of the channel.  (Indeed, using a code that was not strong enough to correct the typical errors might well arouse Eve's suspicions.)  We will also assume, for simplicity of this analysis, that the QECC is {\it nondegenerate}.  This means that on a valid codeword in the QECC, the typical errors $E_{\underline{i}}$ all have distinct error syndromes, and act as unitaries that move the state to a distinct, orthogonal subspace labeled by $\underline{i}$.  This means that error $E_{\underline{i}}$ occurs with a fixed probability $p_{\underline{i}}$ for all valid codewords of the QECC.

We can then essentially repeat the argument that leads to Eq.~(\ref{eq:randomUnitariesRate}), but now using the probabilities $p_{\underline{i}}$.  Note that we now need to take two limits:  the limit of many blocks, and also the limit where the individual blocks are large.  For this argument to apply, we need to first go to the limit of many blocks, and then to the limit of large block size.  In those limits, we can approach a rate
\begin{equation}\label{eq:effectiveEntropy}
- \frac{1}{N} \sum_{\underline{i}} p_{\underline{i}} \log_2 p_{\underline{i}} \equiv \bar{H} ,
\end{equation}
where $\bar{H}$ is an effective entropy per qubit from the channel.

Note that there are some ambiguities in making this argument.  The Kraus map in Eq.~(\ref{eq:onequbitchannel}) is not unique.  Choosing different sets of Kraus operators will lead to different sets of typical errors.  However, these differences should not lead to significant changes to the effective entropy in the limit of large block size, so long as the code is nondegenerate on both sets of typical errors.

\subsection{Secret key consumption}

For the above encodings, how much secret key must be consumed?  In general, we can assume that all the details of the encoding, etc., have been decided between Alice and Bob ahead of time.  So in the protocol as described above, the only place where secret key is consumed is to pick the subsets of errors used in the encoding.

Let's consider the bit flip channel as a simple example.  The possible messages are mapped onto a set of $C$ error syndromes, representing errors of weights $(1-\delta)Np \le w \le (1+\delta)Np$.  For each error weight $w$ in that range, a subset of $C_w$ errors is chosen to represent possible messages.  Alice and Bob can agree before the protocol begins to divide the set of errors of weight $w$ into $n_w$ nonoverlapping subsets of $C_w$ errors each, where
\begin{equation}
n_w = \left(\begin{array}{c} N \\ w \end{array}\right)/C_w
  = \left(\frac{1-p}{p}\right)^{w-Np(1-\delta)} .
\end{equation}
(Since this is unlikely to be an exact integer, one must generally round down, which means that a small fraction of possible errors will be omitted.  This will slightly reduce the match between the steganographic encoding and the noisy channel being simulated, but for large $N$ and $p\ll 1$ the difference will be small.)

For each transmitted block, Alice and Bob must randomly choose one of these $n_w$ subsets for each weight $w$  in the typical range.  Choosing a subset requires $\log_2 n_w$ random bits, which are drawn from their shared key.  However, since any given message is encoded as an error of some specific weight $w$, Alice and Bob can reuse the same secret key bits to choose the subset for each error weight $w$.  So the number of key bits consumed to transmit one block is equal to the maximum value of $\log_2 n_w$ for $(1-\delta)Np \le w \le (1+\delta)Np$, which is 
\begin{eqnarray}
K &=& \max_{Np(1-\delta) \le w \le Np(1+\delta)} \log_2 n_w \nonumber\\
&=& \max_{Np(1-\delta) \le w \le Np(1+\delta)} \log_2 \left(\frac{1-p}{p}\right)^{w-Np(1-\delta)} \nonumber\\
&=& (2Np\delta)  \log_2 \left(\frac{1-p}{p}\right)  .
\end{eqnarray}
How does this scale with $N$?  Since this is a binomial distribution, $\delta$ will take the form
\begin{equation}
\delta = D \sqrt{\frac{1}{N} \left(\frac{1-p}{p}\right) } ,
\end{equation}
where $D$ is a fixed constant determining what fraction of all errors are included in the typical set.  The key consumption therefore is
\begin{equation}
K = 2D \sqrt{N\left(\frac{1-p}{p}\right)} \log_2 \left(\frac{1-p}{p}\right) .
\end{equation}
The key consumption scales sublinearly with $N$, and asymptotically the key consumption rate goes to zero.  While the details will vary, we expect this kind of sublinear scaling of $K$ with $N$ to be generic.  

A few words more on secret key consumption are in order.  In \cite{shaw2011quantum}, Shaw and Brun make a distinction between the {\it secrecy} and the {\it security} of a steganographic protocol.  A steganographic protocol is {\it secret} if an eavesdropper without the secret key cannot distinguish between an encoded message being sent and the noisy channel being applied.  It is {\it secure} if the eavesdropper cannot learn anything about the message, even if she knows that a message is begin sent.

Using a sublinear amount $K$ of shared secret key is sufficient to make the steganographic protocol secret, by this definition.  However, it is {\it not} secure, in general.  Since the number of qubits $M$ transmitted is typically larger than the number of secret key bits $K$ consumed, we would generically expect an eavesdropper to be able to learn on the order of $M-K$ bits of information about the message if she became aware of its existence.

This can be prevented by first encrypting the message before doing the steganographic encoding.  Encryption requires $M$ bits of secret key in the case of a classical message (using a one-time pad), or $2M$ bits of secret key in the case of a quantum message (by twirling).  In this case, the protocol is both secret {\it and} secure.  However, there is a cost:  the secret key is now consumed asymptotically at a linear rate.

\section{Secrecy, Reliability, and Bounds}\label{noiseless}

\subsection{The information processing task}

Here we consider the steganographic scenario as outlined above where Alice is using fake noise to hide her message from Eve, but the actual physical channel she is sending her information over is noiseless. We will consider the task known as {\it entanglement transmission}. This notion of quantum communication encompasses other quantum information-processing tasks such as mixed-state transmission, pure-state transmission, and entanglement generation. We follow closely the discussion of quantum communication in \cite{wilde2013quantum}.

\begin{figure}
\includegraphics[width=0.5\textwidth,height=0.5\textheight,keepaspectratio]{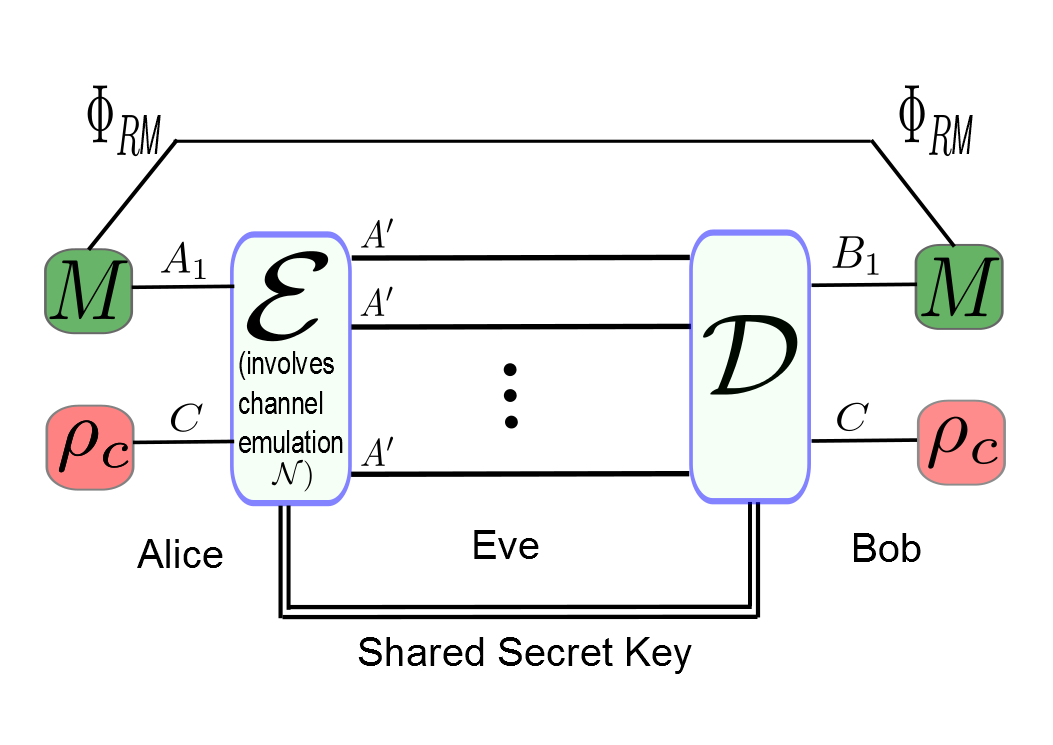}
\caption{The information processing task we consider for Alice sending $M$ stego qubits to Bob over a quantum channel (which is identity for the noiseless case). Alice encodes her message $M$ and an innocent covertext $\rho_c$ into a suitable quantum error-correcting code which has had typical errors applied to it, where the encoding depends on the secret key $k$.  She sends this to Bob, who then decodes the message and covertext using his copy of the shared secret key $k$. Alice's message is entangled with a reference system $R$. The ability to transmit entanglement implies the ability to do general quantum communication.}
\label{fig:protocol}
\end{figure}

The information processing task we are considering is visualized in Figure \ref{fig:protocol}. Alice has a secret message of $M=\log_{2}|A_{1}|$ qubits, which is maximally entangled with a reference system $R$. She also prepares an innocent covertext $\rho_{c}$ which will be encoded into the $N$-qubit quantum error-correcting code. Let us first define her encoded state, dependent on the secret key element $k$:
\begin{equation}
\omega_{k,A'^{n}R}\equiv \mathcal{E}_{k,A_{1}C\rightarrow A^{'n}}(\rho_{c}\otimes\Phi_{A_{1}R}) .
\end{equation}
This dependence of the encoding on the secret key corresponds to choosing among the different sets of error strings $S$ in the protocols from the previous section.  To someone (like Eve) who does not know the secret key $k$, the state is effectively
\begin{equation}
\omega_{A'^{n}R}\equiv\sum_{k} p_k \omega_{k,A'^{n}R} ,
\end{equation}
where $\omega_{A'^{n}R}$ is the state averaged over all possible values of the secret key $k$ with probabilities $p_k$.  (We can choose this probability to be uniform for simplicity, $p_k = p$ for all $k$, if we so desire.)

What is a good way to guarantee secrecy from Eve? We propose the following {\em secrecy} condition:
\begin{equation}\label{eq:stegrlx}
\frac{1}{2}\|\Tr_{R}(\omega_{A'^{n}R})- 
\mathcal{N}^{\otimes N}(V\rho_{c}V^{\dagger})\|_{1}\leq \delta
\end{equation}
where $\mathcal{N}$ is whatever channel Alice is emulating, $V$ is an isometry representing the encoding of the covertext into a suitably chosen codeword (one which can correct typical errors induced by the channel $\mathcal{N}$) and $\delta>0$ is some small parameter. What this condition says is that if Eve observes the quantum state, it will be effectively indistinguishable from an encoded covertext being sent through the noisy quantum channel $\mathcal{N}$.

We introduce another requirement which corresponds to a notion of {\em recoverability}. Once Bob receives the state, he applies his decoder $\mathcal{D}_{k,A'^{n}\rightarrow B_{1}C}$ to obtain the original $\rho_{c}\otimes\Phi_{B_{1}R}$. We can relax this by only requiring that the input states and output states are $\epsilon$ close, that is:
\begin{equation} \label{eq:afw}
\frac{1}{2}\|\mathcal{D}_{k,A'^{n}\rightarrow B_{1}C}(\omega_{k,A'^{n}R})-\rho_{c}\otimes\Phi_{B_{1}R}\|_{1}\leq\epsilon, \forall k
\end{equation}
where $\epsilon > 0$ is a small parameter.

\subsection{Upper bound on steganographic rate}

With these two assumptions of secrecy and recoverability, we can now put a bound on the number of qubits $M$ that can be sent reliably and stegonagraphically from Alice to Bob. Defining $\sigma_{E}\equiv \mathcal{N}^{\otimes N}(V\rho_{c}V^{\dagger})$ and applying the Fannes-Audeneart inequality to the secrecy condition we have:
\begin{equation}\label{eq:secretproof}
H(\Tr_{R}(\omega_{A^{'n}R}))\leq H(\sigma_{E})+\delta N+h_{2}(\delta)
\end{equation}
where $h_{2}$ is the binary entropy function. Furthermore, from the recoverability condition we have
\begin{align}\label{eq:reliableproof}
M&=\log|A_{1}|=I(R\rangle B_{1})_{\Phi} \nonumber \\
&\leq I(R\rangle B_{1})_{\mathcal{D}_{k}(\omega)}+ \epsilon N+(1+\epsilon)h_{2}(\epsilon/[1+\epsilon]) \nonumber \\
&\leq I(R\rangle A'^{n})_{\omega_{k}}+f(N,\epsilon) \nonumber \\
&\leq H(\Tr_{R}(\omega_{k,A^{'n}R}))+f(N,\epsilon).
\end{align}
The first equality follows from the fact that the coherent information of a maximally entangled state is just the logarithm of the dimension of one of the subsystems. The first inequality follows from the AFW inequality applied to \eqref{eq:afw}.  The second inequality is the data processing inequality. The last inequality follows from the definition of the coherent information.

The concavity of entropy implies that
\begin{equation}
\sum_{k} p_k H(\omega_{k,A'^{n}}) \leq H\left(\sum_{k} p_k \omega_{k,A'^{n}}\right)
  = H(\omega_{A'^{n}}) .
\label{eq:concavity}
\end{equation}
The encodings $\mathcal{E}_{k,A_{1}C\rightarrow A^{'n}}$ are isometries, which means that $H(\omega_{k,A'^{n}})$ has the same value for every $k$.  We can therefore sum over the probabilities $p_k$ on the left-hand side of (\ref{eq:concavity}) to get
\begin{equation}
H(\Tr_{R}(\omega_{k,A^{'n}R}))\leq H(\Tr_{R}(\omega_{A^{'n}R})).
\end{equation}

Now putting \eqref{eq:secretproof} and \eqref{eq:reliableproof} together we arrive at our main result, which states that Alice can secretly and reliably send $M$ stego qubits to Bob, where $M$ is bounded above by
\begin{align}\label{eq:bound}
M&\leq H(\Tr_{R}(\omega_{RA^{'n}}))+f(N,\epsilon) \nonumber\\
&\leq H(\sigma_{E})+g(N,\delta)+f(N,\epsilon) ,
\end{align}
where $g(N,\delta)\equiv\delta N+h_{2}(\delta)$. Thus, if we can compute a maximum for $H(\mathcal{N}^{\otimes N}(\rho))$ when $\rho$ is pure (because $V$ is an isometric encoding and $\rho_{c}$ is pure), we have a tight upper bound on the number of qubits $M$ that can be sent steganographically over a noiseless quantum channel. (Of course, if the actual quantum channel is noisy, then this bound will in general be changed.  This is the topic of future work.)

\subsection{Upper bounds for specific channels}

We will now apply our result \eqref{eq:bound} to the channels discussed in the previous section, where we make the implicit assumption that Alice is using a nondegenerate code. Though our result \eqref{eq:bound} is true in general, for a degenerate code the number of distinct error syndromes is smaller (depending on the code), and the bounds discussed here and achievable rates discussed in the previous section would be adjusted.

\subsubsection{The bit flip channel}

For the bit flip channel, i.e., $\mathcal{N}_{BF}(\rho)=(1-p)\rho+pX\rho X$, the maximum of $H(\mathcal{N}^{\otimes N}(\rho))$ over all $N$-qubit pure states $\rho$ is $Nh(p)$ where $h(p)=-p\log p - (1-p)\log (1-p)$ is the entropy of a single qubit sent through a bit flip channel. To prove this, consider some pure state $\rho=\ket{\psi}\bra{\psi}$. Then
\begin{equation}
\mathcal{N}_{BF}^{\otimes N}(\ket{\psi}\bra{\psi})=\sum_{s}p(s)X^{s}\ket{\psi}\bra{\psi}X^{s}
\end{equation}
where we are summing over all binary strings $s$ of length $N$; $X^{s}$ is the operator acting on $N$ qubits with an $X$ acting at every location where $s$ has a 1 and an $I$ where $s$ has a 0. The probability $p(s)$ is given by
\begin{equation}
p(s)= p^{w(s)}(1-p)^{(N-w(s))},
\end{equation}
where $w(s)$ is the weight of string $s$. The Shannon entropy of this distribution is $Nh(p)$, since it is a binomial distribution. The von~Neumann entropy is the minimum Shannon entropy over all possible ensemble decompositions of the given state, and it is not hard to check that it is achieved when $\ket{\psi}$ is a $Z$ eigenstate. Thus the encoding described in the previous section for steganography with an simulated bit flip channel is essentially optimal.

\subsubsection{More general channels}

Unfortunately, for a more general quantum channel $\mathcal{N}$ we may not know, in general, what $N$-qubit pure state $\rho$ maximizes $H(\mathcal{N}^{\otimes N}(\rho))$. However, we can still bound this quantity. First, consider a general quantum channel $\mathcal{N}$ that acts on an $N$ qubit pure state as follows:
\begin{equation}
\mathcal{N}^{\otimes N}(\rho)\approx\sum_{j}E_{j}\rho E_{j}^{\dagger}
\end{equation}
where $\{E_{j}\}$ is the set of typical errors associated with $N$ applications of the channel $\mathcal{N}$. Recall that we are choosing our isometric encoding to correct for typical errors of whatever channel $\mathcal{N}$ it is we are emulating. Though the set of correctable errors $\{E_{j}\}$ need not act like unitaries on the codespace, we  can always find a set of correctable errors $\{\widetilde{E}_{j}\}_{j}$ that do \cite{nielsen2002quantum}. To see this, first consider the Knill-Laflamme condition:
\begin{equation}
\mathbb{P}E_{i}^{\dagger}E_{j}\mathbb{P}=\alpha_{ij}\mathbb{P}
\end{equation}
where $\mathbb{P}$ is the codespace projector and $\alpha$ is a Hermitian matrix. Thus, we can write $\widetilde{\alpha}=U^{\dagger}\alpha U$ where $U$ is a unitary matrix and $\widetilde{\alpha}$ is diagonal.
\begin{equation}
\widetilde{E}_{k}=\sum_{j}M_{jk}E_{k}
\end{equation}
where the unitary $M$ is chosen in such a way as to diagonalize $\alpha$. That is
\begin{align}
\mathbb{P}\widetilde{E}_{k}^{\dagger}\widetilde{E}_{l}\mathbb{P}&=\sum_{ij}M_{ik}^{*}M_{jl}\mathbb{P}E_{i}^{\dagger}E_{j}\mathbb{P} \nonumber \\
&= (\sum_{ij}M_{ik}^{*}\alpha_{ij}M_{jl})\mathbb{P} \nonumber \\
&=\widetilde{\alpha}_{kl}\mathbb{P} = \delta_{kl}\widetilde{\alpha}_{kk}\mathbb{P} .
\end{align}
Note that these errors $\{\widetilde{E}_{j}\}$ act unitarily on the codespace. So long as the Knill-Laflamme condition is satisfied, we can always diagonalize $\alpha$ in this way. Now going back to our expression for the channel action,
\begin{equation}
\sum_{j}E_{j}\rho E_{j}^{\dagger} = \sum_{k,l,j}M_{kj}M^{*}_{lj}\widetilde{E}_{k}\rho\widetilde{E}_{l}^{\dagger} = \sum_{k} \widetilde{E}_{k}\rho\widetilde{E}_{k}^{\dagger}.
\end{equation}
Because we have assumed that the typical errors are all correctable, and that the code is nondegenerate, the states $\widetilde{E}_{k}\rho\widetilde{E}_{k}^{\dagger}$ are all orthogonal to each other, and $\Tr\{\widetilde{E}_{k}\rho\widetilde{E}_{k}^{\dagger}\} = \alpha_{kk}$.  The von~Neumann entropy is the Shannon entropy minimized over all possible decompositions, so the entropy of this state is clearly
\begin{equation}
H(\sigma_{E})=H(\mathcal{N}^{\otimes N}(V\rho_{c}V^{\dagger})) \leq -\sum_{k}\alpha_{kk}\log_{2}(\alpha_{kk}) .
\end{equation}
By \eqref{eq:bound} we have shown that the amount of steganographic communication allowed for a quantum channel $\mathcal{N}$ emulation is upper bounded by this quantity. Applying this to the general channel discussed in section II.C above, we see that this quantity is equal to $N\bar{H}$, where $\bar{H}$ is the effective entropy per qubit defined in Eq.~(\ref{eq:effectiveEntropy}).  So this encoding approaches the maximum possible rate for the general channel, just as for the bit flip channel.

\section{Conclusions and Future Work}\label{conclusion}

Quantum steganography is the study of secret quantum communication between two parties, Alice and Bob. We have shown that Alice and Bob are able to communicate with each other secretly at a nonzero rate using a shared secret key, without arousing suspicion from a potential eavesdropper Eve. In this paper we gave explicit bounds on the number of stego qubits that Alice can send to Bob when Alice is simulating a general quantum channel $\mathcal{N}$ with her stego encoded message, as well as explicit encodings to that achieve these bounds, for the case when the actual physical channel is noiseless.

The obvious next question is what if the channel shared between Alice and Bob (as is generally the case) is noisy? There is reason to believe that so long as Eve has some ignorance about the actual physical channel, then Alice will still be able to communicate steganographically to Bob.

For instance, suppose the actual physical channel is a depolarizing channel $\mathcal{N}_{p}$ where $p$ is the depolarizing parameter and the channel that Eve expects is $\mathcal{N}_{p+\epsilon-4p\epsilon/3}$ for some small suitably chosen $\epsilon>0$. Then Alice can emulate a depolarizing channel $\mathcal{N}_{\epsilon}$ in such a way such that if Eve observes the state Alice is sending to Bob, it will look like an innocent encoded covertext passing through $N$ applications of a channel $\mathcal{N}_{p}\circ\mathcal{N}_{\epsilon}$ (where $N$ is the length of the codeword Alice is using). There should be elements of the encoding given in this paper that will generalize to the noisy case for general channels $\mathcal{N}$. This will certainly be an area of fruitful future study.

\section*{Acknowledgments}
Thanks to Mark Wilde and David Ding for helpful discussions.  This research was supported in part by  NSF Grants CCF-1421078 and QIS-1719778, and by an IBM Einstein Fellowship at the Institute for Advanced Study.
\bibliography{qstegnoisebib}

\end{document}